\documentclass[a4paper,twoside]{article}

\usepackage{epsfig}
\usepackage{subcaption}
\usepackage{calc}
\usepackage{amssymb}
\usepackage{amstext}
\usepackage{amsmath}
\usepackage{amsthm}
\usepackage{multicol}
\usepackage{pslatex}
\usepackage{apalike}
\usepackage{algorithm2e}
\usepackage[bottom]{footmisc}
\usepackage{amsfonts}
\usepackage{algorithmic}
\usepackage{graphicx}
\usepackage{textcomp}
\usepackage{xcolor}
\usepackage{physics}
\usepackage{booktabs}
\usepackage{braket}
\usepackage{microtype}
\usepackage[hidelinks]{hyperref}
\usepackage[capitalize,noabbrev]{cleveref}
\usepackage{array}
\usepackage{multirow}
\usepackage{stfloats}  

\usepackage{SCITEPRESS}     

\graphicspath{ {./figures/} }

\begin{document}

\title{Anomaly Detection with Quantum SVR in the NISQ Era: Limits of Robustness to Noise and Adversarial Attacks}

\author{\authorname{Kilian Tscharke\orcidAuthor{0009-0006-7423-2498}, Maximilian Wendlinger\orcidAuthor{0009-0008-1031-7665}, Sebastian Issel\orcidAuthor{0009-0002-3873-1979} and Pascal Debus\orcidAuthor{0009-0007-2437-2764}}
\affiliation{Fraunhofer Institute for Applied and Integrated Security (AISEC), Garching near Munich, Germany}
\email{\{firstname.lastname\}@aisec.fraunhofer.de}
}


\abstract{Anomaly Detection (AD) is critical in data analysis, particularly within the domain of IT security.
In this study, we explore the potential of Quantum Machine Learning for application to AD with special focus on the robustness to noise and adversarial attacks. 
We build upon previous work on Quantum Support Vector Regression (QSVR) for semisupervised AD by conducting a comprehensive benchmark on IBM quantum hardware using eleven datasets. Our results demonstrate that QSVR achieves strong classification performance and even outperforms the noiseless simulation on two of these datasets.
Moreover, we investigate the influence of -- in the NISQ-era inevitable -- quantum noise on the performance of the QSVR. Our findings reveal that the model exhibits robustness to depolarizing, phase damping, phase flip, and bit flip noise, while amplitude damping and miscalibration noise prove to be more disruptive.
Finally, we explore the domain of Quantum Adversarial Machine Learning by demonstrating that QSVR is highly vulnerable to adversarial attacks, with neither quantum noise nor adversarial training improving the model’s  robustness against such attacks.}

\onecolumn \maketitle \normalsize \setcounter{footnote}{0} \vfill

\section{\uppercase{Introduction}}
\label{sec:introduction}

Quantum Kernel Methods (QKMs) have gained attention in Quantum Machine Learning (QML) as they can replace many supervised QML models \cite{supervised_qml_are_kernel_schuld2021}.
Theoretically, QKMs offer a promising route to tackle classification problems that are intractable for classical methods. In particular, Support Vector Machines (SVMs) with quantum kernels (QSVM) have been shown to achieve exponential speedups on certain tasks, such as classification based on the discrete logarithm problem~\cite{quantum_kernel_advantage_discrete_log_2021} or the k-Forrelation problem \cite{J_ger_2023}.

Anomaly Detection (AD) is crucial in IT security, where it detects deviations from normal behavior in domains like intrusion and fraud detection \cite{Ruff2020}. However, ML models are vulnerable to adversarial attacks \cite{adv_attacks_szegedy} -- small, crafted perturbations that cause misclassification -- studied in Adversarial Machine Learning (AML).

Given the potential of QML to address problems challenging for classical methods, the application of QML -- and especially QKMs -- to AD is a tempting progression.
\cite{Tscharke} proposed a semisupervised AD approach based on the reconstruction loss of a Quantum Support Vector Regression (QSVR) equipped with a quantum kernel. 
Their choice of QSVR was motivated by its close connection to the QSVM, which has demonstrated theoretical advantages over classical methods in specific problem settings. Moreover, the regression formulation offers a natural way to produce a continuous anomaly score, providing a quantitative measure of how strongly an input deviates from expected normal behavior.
The authors compared the performance of the QSVR against classical and quantum baselines on eleven datasets. Their simulated QSVR demonstrated comparable performance to a classical SVR, with marginal superiority over the other models. 
However, their work did not address implementation on quantum hardware, nor did it investigate the adversarial robustness of the QSVR, two gaps this paper tackles to assess the model's reliability in security-critical applications.

In the current Noisy Intermediate-Scale Quantum (NISQ)-era, hardware limitations such as decoherence and gate errors pose significant challenges for deploying quantum algorithms in practical settings. These noise effects not only degrade the performance of QML models but also influence their susceptibility to adversarial attacks. As shown in studies~\cite{quantum_noise_against_adv_attacks_Du,noise_improve_adv_robustness_Huang}, noise can both undermine and, paradoxically, enhance model robustness depending on its nature and interaction dynamics. This dual role underscores the importance of investigating how noise impacts both the classification performance and adversarial resilience of quantum models like the QSVR, two aspects that this paper examines.

The remainder of this work is structured as follows. Section~\ref{Related Work} and Section~\ref{Background} review related work and provide the necessary preliminaries. Section~\ref{Methods} describes the experimental setup, Section~\ref{Results and Discussion} presents and discusses the results, and Section~\ref{Conclusion and Outlook} concludes with future directions.

\subsection{Related Work} \label{Related Work}
\cite{Havlicek2019} introduced a QSVM for binary classification on two qubits of a five-qubit NISQ device.
Since then, QSVMs have been applied to many areas, including remote sensing image classification \cite{Delilbasic2021}, mental health treatment prediction \cite{Ahmad2021}, and breast cancer prediction \cite{Mafu2021}. 
\cite{quantum_one_class_SVM} extended this to unsupervised fraud detection with a simulated one-class QSVM, and \cite{Tscharke} developed a QSVR for semisupervised AD in 2023.
However, a QSVR for semisupervised AD has not yet been set up on hardware.

Research has also explored the influence of noise on QML models \cite{noise_influence_Diego,noise_influence_winderl}, focusing on noise robustness \cite{noise_robustness_Nguyen,noise_robustness_Yao} or beneficial use of noise \cite{noise_beneficial_ju,noise_beneficial_winderl}. However, the influence of noise on a QSVR for semisupervised AD has not yet been evaluated.

Finally, the link between quantum noise and QAML was established by \cite{quantum_noise_against_adv_attacks_Du}, who found that adding depolarization noise can increase adversarial robustness. Building on this, \cite{noise_improve_adv_robustness_Huang} improved the adversarial robustness of Quantum Neural Networks by adding noise layers. To date, there have been no published results involving adversarial attacks on semisupervised QSVR for AD.

\subsection{Contributions} \label{Contributions}
The goal of our work is to gain further insight into the potential of QSVR for semisupervised AD in the NISQ-era, which we accomplish through these contributions: 
\begin{enumerate}
    \item We investigate how the model performs on hardware and report that the QSVR achieves good classification performance on five qubits of a 27-qubit IBM device, even outperforming a noiseless simulation on two out of eleven datasets.
    \item We show that the QSVR is largely robust to noise by simulating over 500 noisy models. We further observe that \emph{amplitude damping} and \emph{miscalibration} have the most damaging effect on the model's performance and that the artificial \emph{Toy} dataset, constructed to be linearly separable, suffers the most from noise.
    \item Finally, we investigate the robustness of the QSVR against adversarial attacks and find it highly vulnerable, with performance on real-world datasets dropping by up to an order of magnitude for a weak attack strength of $\varepsilon=0.01$. Introducing noise into the QSVR or carrying out adversarial training does not clearly improve the model's adversarial robustness.
\end{enumerate}

\section{\uppercase{Background}} \label{Background}

 \subsection{Quantum Kernel Methods}
 A kernel is a positive, semidefinite function $\kappa: \mathcal{X} \times \mathcal{X} \to \mathbb{R}$  on the input set $\mathcal{X}$. It uses a similarity measure $\kappa (\boldsymbol{x}_i, \boldsymbol{x}_j)$ between two input vectors $\boldsymbol{x}_i, \boldsymbol{x}_j \in \mathcal{X}$ to create a model that captures the properties of a data distribution. A \emph{feature map} $\phi : \mathcal{X} \to \mathcal{F}$ maps input vectors $\boldsymbol{x}$ to a Hilbert or \emph{feature space} $\mathcal{F}$. They are of great importance in ML, as they map input data in a higher-dimensional space. The feature map can be a nonlinear function that changes the relative position of the data points. As a result, the dataset can become easier to classify in feature space, and even linearly separable. We associate feature maps with kernels by defining a kernel via
 \begin{align}
     \kappa (\boldsymbol{x}_i, \boldsymbol{x}_j) := \braket{\phi(\boldsymbol{x}_i), \phi(\boldsymbol{x}_j)}_\mathcal{F}
 \end{align}
where $\braket{\cdot, \cdot}_\mathcal{F}$ is the inner product defined on $\mathcal{F}$.

With the exponentially large Hilbert space of QCs, the use of QKMs for ML is close at hand. A quantum feature map $\psi: \boldsymbol{x} \to \ket{\phi(\boldsymbol{x})}$ is implemented via a feature-embedding circuit $U(\boldsymbol{x})$, which acts on a ground state $\ket{0 \dots 0}$ of a Hilbert space $\mathcal{F}$ as $\ket{\phi(\boldsymbol{x})} = U(\boldsymbol{x}) \ket{0 \dots 0}$. 
The distance measure in the quantum kernel is the absolute square of the inner product of the quantum states.
On hardware, this can be realized by the \emph{inversion test}, where a sample $\boldsymbol{x}_i$ is encoded in the unitary $U$, followed by the adjoint $U^\dagger$ encoding the second sample $\boldsymbol{x}_j$ and measuring the probability of the all-zero state. Thus, the quantum kernel is defined as
\begin{align}
    \kappa(x_i, x_j) &= \left|\braket{\phi(\boldsymbol{x}_i) | \phi(\boldsymbol{x}_j)}\right|^2 \notag \\
    &= \left| \bra{0^{\otimes n}} U^\dagger(\boldsymbol{x}_i) U(\boldsymbol{x}_j) \ket{0^{\otimes n}} \right| ^2
\end{align}
and returns the \emph{overlap} or \emph{fidelity} of the two states. A more in-depth description of quantum kernels can be found in \cite{Schuld_ML_w_QC,Schuld_QML_in_Feature_Hilbert_Spaces}.

There exist many different encoding techniques for the circuit realizing the quantum feature map, but for this work, we will focus on \emph{angle encoding} because of its advantageous complexity with respect to the number of gates  $\mathcal{O}(k)$ for an input vector $\textbf{x}$ of dimension $k$.
\emph{Angle encoding} is a special form of time-evolution encoding, where a scalar value $\boldsymbol{x}$ is encoded in the unitary evolution of a quantum system governed by a Hamiltonian $H$. The unitary of time-evolution encoding is given by
\begin{align}
    U(x) = e^{-i\boldsymbol{x}H}.
\end{align}
In the case of \emph{angle encoding}, the Pauli matrices $\sigma_a$ with $a \in \{x, y, z\}$ are used in the Hamiltonian $H = \frac{1}{2} \sigma_a$. \\

 \subsection{Noisy Quantum Computing} \label{Noisy Quantum Computing}
Real quantum systems are never completely isolated from the environment; for example, an electron realizing a qubit will interact with other charged particles. Moreover, quantum computers are programmed by an external system and thus can never be a fully closed system \cite{Nielsen,Suter_Noise}.
 
In the current NISQ-era, noise significantly limits the performance of quantum algorithms, primarily through \emph{coherent} and \emph{incoherent} noise. \emph{Coherent} noise arises from systematic, reversible errors that lead to predictable but undesired evolution of the states, often caused by imperfect calibrations or imprecise control signals \cite{kaufmann2023_noise}. It is a \emph{unitary} evolution of the system, described by only one operation element in the \emph{operator-sum representation} introduced below \cite{preskill_lectures_noise,Wallman_Noise}.

 \emph{Incoherent} noise, on the other hand, rises from random, stochastic processes caused by insufficient isolation of the system from its environment. These uncontrolled interactions between system and environment lead to deviations between the desired and the actual evolution of the system and to a loss of coherence \cite{kaufmann2023_noise,Suter_Noise}. 
 
 Quantum noise can be modeled by a quantum channel, where the term "channel" is drawn from classical information theory \cite{Nielsen}. In the \emph{operator-sum representation}, a channel is described by the map $\mathcal{E}$ with operation elements (or \emph{Kraus operators}) $\{E_i\}$ mapping the density operator $\rho = \ket{\psi} \bra{\psi}$ to another density operator $\mathcal{E}(\rho)$.
 \begin{align}
    \mathcal{E}(\rho) = \sum_i E_i \rho E_i^\dagger .
 \end{align}
 In the following, selected types of single-qubit noisy channels are described, and their operation elements $E_i$ are listed. For a more detailed explanation of the noisy quantum channels, we refer to \cite{Nielsen,preskill_lectures_noise}. 
 
 \begin{enumerate}
\item \emph{Amplitude Damping} channel: describes the effect of energy loss, such as when an atom emits a photon. The channel acts on the quantum system $A$ and the environment $E$ as follows: if both  $A$ and $E$ are in their ground state $\ket{0}$, nothing happens. If $A$ is in the excited state $\ket{1}_A$, a photon will be emitted with probability $p$, leading to the excitation of the environment and causing the transition $\ket{0}_E \to \ket{1}_E$, while $A$ drops to the ground state, i.e. $\ket{1}_A \to \ket{0}_A$. The evolution caused by the channel can be summarized as:
\begin{align}
& |0\rangle_A \otimes|0\rangle_E \mapsto|0\rangle_A \otimes|0\rangle_E  \\
& |1\rangle_A \otimes|0\rangle_E \mapsto \sqrt{1-p}|1\rangle_A \otimes|0\rangle_E+\sqrt{p}|0\rangle_A \otimes|1\rangle_E \notag
\end{align}

This is achieved by the operation elements:
 \begin{align}
     E_0 &= \left[\begin{array}{rr}
1 & 0 \\
0 & \sqrt{1-p}
\end{array}\right] \notag \\
E_1 &= \left[\begin{array}{rr}
0 & \sqrt{p} \\
0 & 0
\end{array}\right]
 \end{align}

\item \emph{Bit Flip} channel: flips the state of a qubit from $\ket{0}$ to $\ket{1}$ and vice versa with probability $p$. The operators are:
 \begin{align}
     E_0=\sqrt{1-p} I=\sqrt{1-p}\left[\begin{array}{rr}
1 & 0 \\
0 & 1
\end{array}\right] \notag \\
E_1=\sqrt{p} X=\sqrt{p}\left[\begin{array}{rr}
0 & 1 \\
1 & 0
\end{array}\right]
 \end{align}

\item \emph{Depolarizing} channel: the qubit remains intact with probability $1-p$, while an error occurs with probability $p$. If an error occurs, the state is replaced by a uniform ensemble of the three states $X \ket{\psi},Y \ket{\psi}, Z \ket{\psi}$. This is a symmetric decoherence channel defined by operation elements:
 \begin{align}
     E_0 &=\sqrt{1-p} I &=\sqrt{1-p}\left[\begin{array}{rr}
1 & 0 \\
0 & 1
\end{array}\right] \notag \\
X\text{-Error: } E_1 &= \sqrt{p/3} X &= \sqrt{p/3}\left[\begin{array}{rr}
0 & 1 \\
1 & 0
\end{array}\right] \notag \\
Y\text{-Error: } E_2 &= \sqrt{p/3} Y &=\sqrt{p/3}\left[\begin{array}{rr}
0 & -i \\
i & 0
\end{array}\right] \notag \\
Z\text{-Error: } E_3 &= \sqrt{p/3} Z &=\sqrt{p/3}\left[\begin{array}{rr}
1 & 0 \\
0 & -1
\end{array}\right] 
 \end{align}

\item \emph{Miscalibration} channel: a coherent noise channel applying an "overrotation"  $p$ to the $R_a$ rotation gate with $a \in \{x,y,z\}$. This can be caused by an imperfect calibration of the device \cite{kaufmann2023_noise}. Since the channel is unitary, it has only one operation element:
 \begin{align}
     E_0 = R_a(p).
 \end{align}
 
\item \emph{Phase Damping} channel: the \emph{phase damping} or \emph{dephasing} channel
models the effect of random environmental scattering on a qubit, such as photon interactions in a waveguide or atomic states perturbed by distant charges.
This channel causes a partial loss of phase information without energy loss. It produces the same effect as the \emph{phase flip} channel, with the phase damping  $\lambda$ related to the phase flip probability $p$ by
\begin{align}
    p = \frac{1-\sqrt{1-\lambda}}{2} .
\end{align} 

\item \emph{Phase Flip} channel: applies a phase of $-1$ to the $\ket{1}$-state with probability $p$ and leaves the $\ket{0}$-state unchanged. 
It has the operation elements:
 \begin{align}
     E_0=\sqrt{1-p} I=\sqrt{1-p}\left[\begin{array}{rr}
1 & 0 \\
0 & 1
\end{array}\right] \notag \\
E_1=\sqrt{p} Z=\sqrt{p}\left[\begin{array}{rr}
1 & 0 \\
0 & -1
\end{array}\right]
 \end{align}

The channel has the following effect on the state's density matrix $\rho$:
\begin{align}
\mathcal{E}\left(\begin{array}{ll}
\rho_{00} & \rho_{01} \\
\rho_{10} & \rho_{11}
\end{array}\right)=\left(\begin{array}{cc}
\rho_{00} & (1-2p) \rho_{01} \\
(1-2p) \rho_{10} & \rho_{11}
\end{array}\right)
\end{align}
From this, we can see that the phase flip channel destroys superposition by decaying the off-diagonal terms of the density matrix $\rho$ while the on-diagonal terms remain invariant.  
\end{enumerate}

\subsection{Adversarial Attacks}
QML models are typically trained using a hybrid quantum-classical approach. In this framework, the model parameters are optimized using a classical optimization algorithm, while the quantum part is limited to the evaluation of the loss, which is (partially) done by the quantum computer or simulator. This hybrid approach allows us to easily extend the concept of adversarial attacks from classical ML to QML, which has already been successfully demonstrated in \cite{adv_robustness_wendlinger}.

An adversarial attack is a small, carefully crafted perturbation of the $k$-dimensional input $\boldsymbol{x}$ that causes the model to misclassify the input \cite{adv_attacks_szegedy,adv_attacks_goodfellow}. An untargeted adversarial sample is created by maximizing the model's loss $\mathcal{L}$ while keeping the perturbation $\delta$ small enough to be imperceptible to humans, e.g., by ensuring $\delta \in \Delta=\{\delta \in \mathbb{R}^{k}: \| \delta\|_{\infty} \le \varepsilon\}$ for some small $\varepsilon$. In general, the ideal perturbation is given by
\begin{align}
\delta \equiv \underset{\delta^{\prime} \in \Delta}{\operatorname{argmax}} \; \mathcal{L}\left(f\left(x+\delta^{\prime} ; \theta^*\right), y\right),
\end{align}
where $f$ is the model, $\theta^*$ are the model's optimized parameters after training, and $y$ is the target.

One of the most widely used methods for generating adversarial samples is Projected Gradient Descent (PGD) \cite{pgd_madry}. PGD iteratively maximizes the model's prediction error while ensuring that the perturbation remains within a predefined range. This approach has become a standard tool for evaluating the robustness of models to adversarial threats. The perturbed input is determined by
\begin{align}
\boldsymbol{x}^{t+1}=\Pi_{\boldsymbol{x}+S}\left(\boldsymbol{x}^t + \alpha \operatorname{sgn}\left(\nabla_{\boldsymbol{x}} \mathcal{L}(\boldsymbol{x}^t, y; \theta^*)\right) \right),
\end{align}

where $\boldsymbol{x}^t$ represents the perturbed data at step $t$, $\Pi_{\boldsymbol{x}+S}$ clips the perturbed data into the range of the normalized input set $S$, and $\alpha$ is the step size.

A straightforward strategy to increase the adversarial robustness is \emph{adversarial training}, where adversarial samples are included in the training set.

\section{\uppercase{Methods}} \label{Methods}
The experiments performed in this work are threefold. First, we benchmarked the QSVR for semi-supervised AD on hardware. Second, we evaluated the influence of noise on the classification performance of the QSVR, and third, we investigated the influence of noise and adversarial retraining on the adversarial robustness of the QSVR.
An overview of the datasets used in the experiments is given in \cref{app:tab:overview_datasets} in the appendix. The datasets were reduced to five dimensions using Principal Components Analysis.
For a more detailed description of the model, the datasets, and the preprocessing techniques, we refer to \cite{Tscharke}.

\subsection{Quantum Support Vector Regression Model}
\begin{figure*}[htbp]
\centerline{\includegraphics[width=\textwidth]{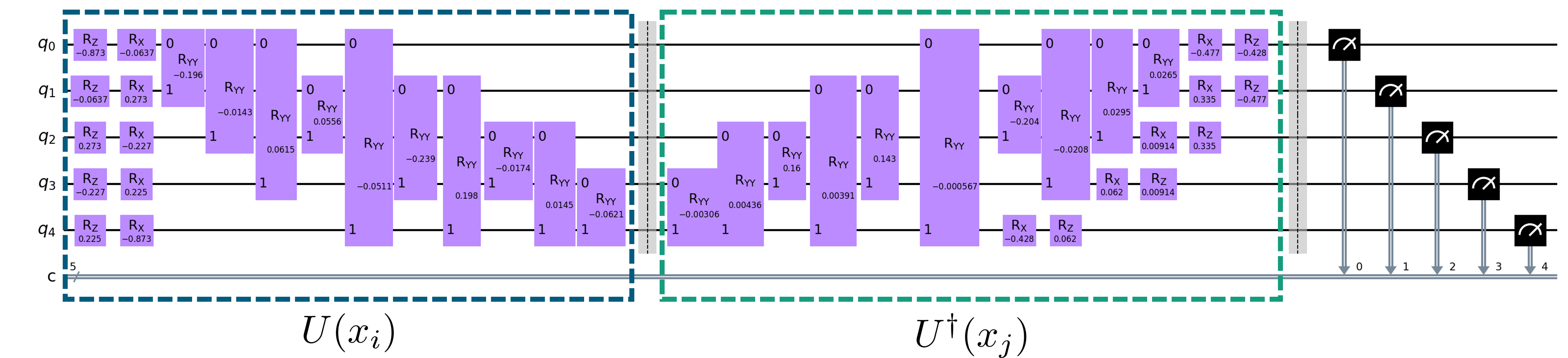}}
\caption{QSVR circuit. Explanation in text.}
\label{fig_QSVR_circuit}
\end{figure*}

The QSVR architecture for semi-supervised AD was originally introduced in \cite{Tscharke}, and the corresponding kernel circuit is illustrated in \cref{fig_QSVR_circuit}. To compute a kernel entry, the data point $\boldsymbol{x}_i$ is first encoded by the unitary $U$, followed by the application of its inverse $U^\dagger$ embedding a second data point $\boldsymbol{x}_j$. Our $U$ consists of a layer of RZ and RX gates, followed by IsingZZ\footnote{\url{https://docs.pennylane.ai/en/stable/code/api/pennylane.IsingZZ.html}} gates that generate entanglement. Each single-qubit rotation encodes one feature, while the parameter of each IsingZZ gate is given by the product of two features. The kernel value $K_{ij} = K_{ji}$ is obtained by measuring the probability of the all-zero state after execution of $U^\dagger U$.

\subsection{Hardware Experiments}
In the hardware experiments, we used a training set of size 30 from the normal class and a test set of size 50 with equal class ratio, following \cite{Tscharke}.
\subsubsection{Device Specifications}
The experiments were performed on the IBM System One in Ehningen, Germany. The system is a 27-qubit superconducting quantum computer with a quantum volume of 64. The QSVR was implemented using qiskit \cite{qiskit} with  default error mitigation techniques. Further specifications of the system are listed in table~\ref{tab_Ehningen_Specs}.

\begin{table}[htbp]
\centering
\caption{Specification of the IBM System One Ehningen.}
\label{tab_Ehningen_Specs}
\begin{tabular}{ll}
\toprule 
Spec & Value\\
\midrule 
Name & IBM Quantum Sys- \\
     & tem One at Ehningen \\
System type & Superconducting \\
Number of qubits & 27\\
Quantum Volume & 64 \\
Processor type & Falcon \\
Deployment year & 2021 \\
Coherence time & $\approx 150\mu$s \\
Single qubit error & $\approx 0.025\%$ \\
Two qubit gate error & $\approx 0.7\%$ \\
Operation time of CNOT & $\approx 300$ns \\
\bottomrule 
\end{tabular}
\end{table}

 \subsubsection{Reference Models}
 We benchmarked our model against four other models, following the approach in \cite{Tscharke}. The models include a simulated QSVR, a simulated quantum autoencoder based on \cite{Kottmann2021}, a classical SVR, and a classical autoencoder.

\subsection{Generation of Noise} \label{Methods Generation of Noise}
The influence of noise on the QSVR for semisupervised AD was evaluated by applying six noise channels with different strengths to the quantum circuit that computes the kernel. The seven noise probabilities used in the experiments are $p \in \{0.01, 0.05, 0.1, 0.2, 0.3, 0.4, 0.5\}$ and the five incoherent noise channels are [\emph{bit flip}, \emph{phase flip}, \emph{depolarizing}, \emph{phase damping}, \emph{amplitude damping}]. This leads to a total of $7 \cdot 5 = 35$ models per dataset.
For the \emph{miscalibration} channel, the noise probability $p$ is the overrotation in radians in 20 linear steps between 0 and $2\pi$. For the DoH dataset subject to adversarial attacks of strength $\varepsilon=0.1$, additional evaluations were performed in the region close to $p=\pi$.
The noisy QSVR was simulated using PennyLane \cite{pennylane}. We used a training set of size 100 from the normal class and a test set of size 100 with a balanced class ratio.

\subsection{Generation of Adversarial Attacks}
We created 100 adversarial samples of the test set (50 from each class) using PGD with the parameters listed in Table \ref{tab:adv_attacks}. The attacks targeted the noiseless models and were then applied to the noisy models. For adversarial training, we create adversarial samples of the training set and train the model using the adversarial training set of size 100.

\begin{table}[htbp]
\centering
\caption{Overview of the parameters used in the PGD attacks.}  \label{tab:adv_attacks}
\begin{tabular}{ll}
\toprule 
Spec & Values\\
\midrule 
Attack strength $\varepsilon$ & [0.01, 0.1, 0.3] \\
Iterations $n$ & 50 \\
$\alpha$ & $\varepsilon/n$ \\
\bottomrule 
\end{tabular}
\end{table}

\section{\uppercase{Results and Discussion}} \label{Results and Discussion}
In this study, we first benchmarked our QSVR on the 27-qubit IBM Ehningen device (labeled \emph{qc-QSVR}) and compared its performance against the simulated quantum baseline models QSVR (simulated version of our model, see \cite{Tscharke}) and QAE (based on \cite{Kottmann2021}), as well as CSVR and CAE as classical baselines.
Second, six different noise channels of varying strength were introduced to evaluate the influence of noise on the QSVR algorithm. 
Third, the adversarial robustness of the model was examined, and the influence of noise on the adversarial robustness was evaluated by exposing the (noisy) model to adversarial attacks.
The simulations show no error bars because the SVR is a deterministic model and pennylane's \textit{default.mixed}\footnote{\url{https://docs.pennylane.ai/en/stable/code/api/pennylane.devices.default_mixed.DefaultMixed.html}} device used to calculate the kernels computes exact outputs.

\subsection{Hardware Results}
\begin{figure}[htbp]
\centerline{\includegraphics[width=0.5\textwidth]{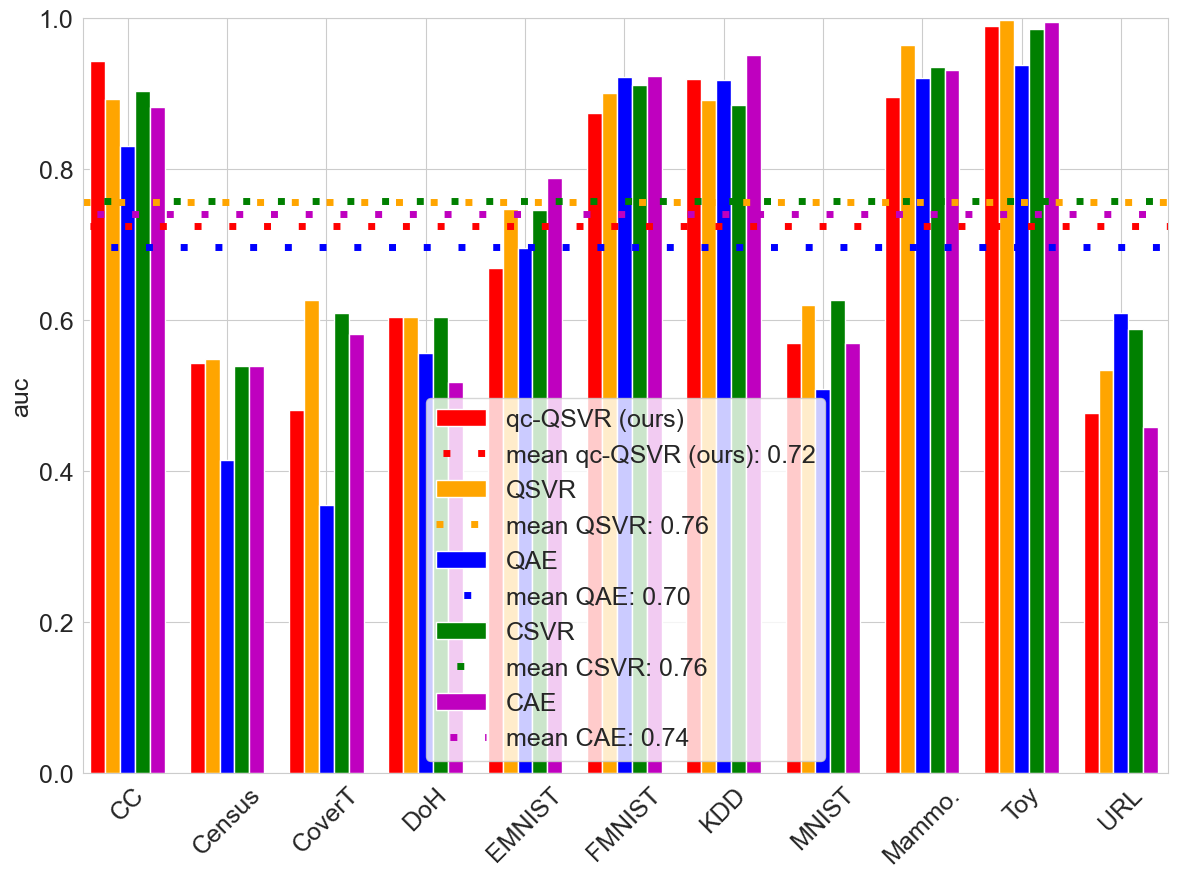}}
\caption{Plot of area under the curve (AUC) for the different models on the evaluated data sets.}
\label{fig_all_auc_bar}
\end{figure}

Figure \ref{fig_all_auc_bar} compares model performance on eleven datasets using area under the ROC curve (AUC).
The datasets include Credit Card Fraud (CC), Census, Forest Cover Type (CoverT), Domain Name System over HTTPS (DoH), EMNIST, Fashion MNIST (FMNIST), Network Intrusion (KDD), MNIST, Mammography (Mammo), URL, and our constructed dataset Toy. 
The models included in the study are qc-QSVR (ours), QSVR, QAE, CSVR, and CAE. The average performance of each model across all datasets is represented by a dotted line, while the bars indicate the models' performance on individual datasets.

On average, our qc-QSVR exhibits an AUC decrease of 0.04 compared to the simulated QSVR. In 8 out of 11 datasets, the hardware model is outperformed by the simulated version, with the performance gap attributed to hardware noise. On the DoH datasets, both models perform identically, while on the CC and KDD datasets, the qc-QSVR surprisingly outperforms its simulated counterpart.
On these two datasets, hardware-induced noise appears to enhance model performance, an effect we ascribe to improved generalization. Specifically, the perturbations introduced by the noisy gates might mimic the corruptions applied in denoising autoencoders, a technique shown to yield superior generalization compared to standard autoencoders \cite{denoising_autoencoder}. 
These results are consistent with those of other authors showing that noise can, under certain conditions, improve the performance of QML models \cite{Escudero_2023,Domingo2023}.
This phenomenon of superior model performance caused by hardware noise will be the subject of future research.

\subsection{Influence of Noise on the Model Performance}
\begin{figure*}[tb]
\centerline{\includegraphics[width=1.097\textwidth]{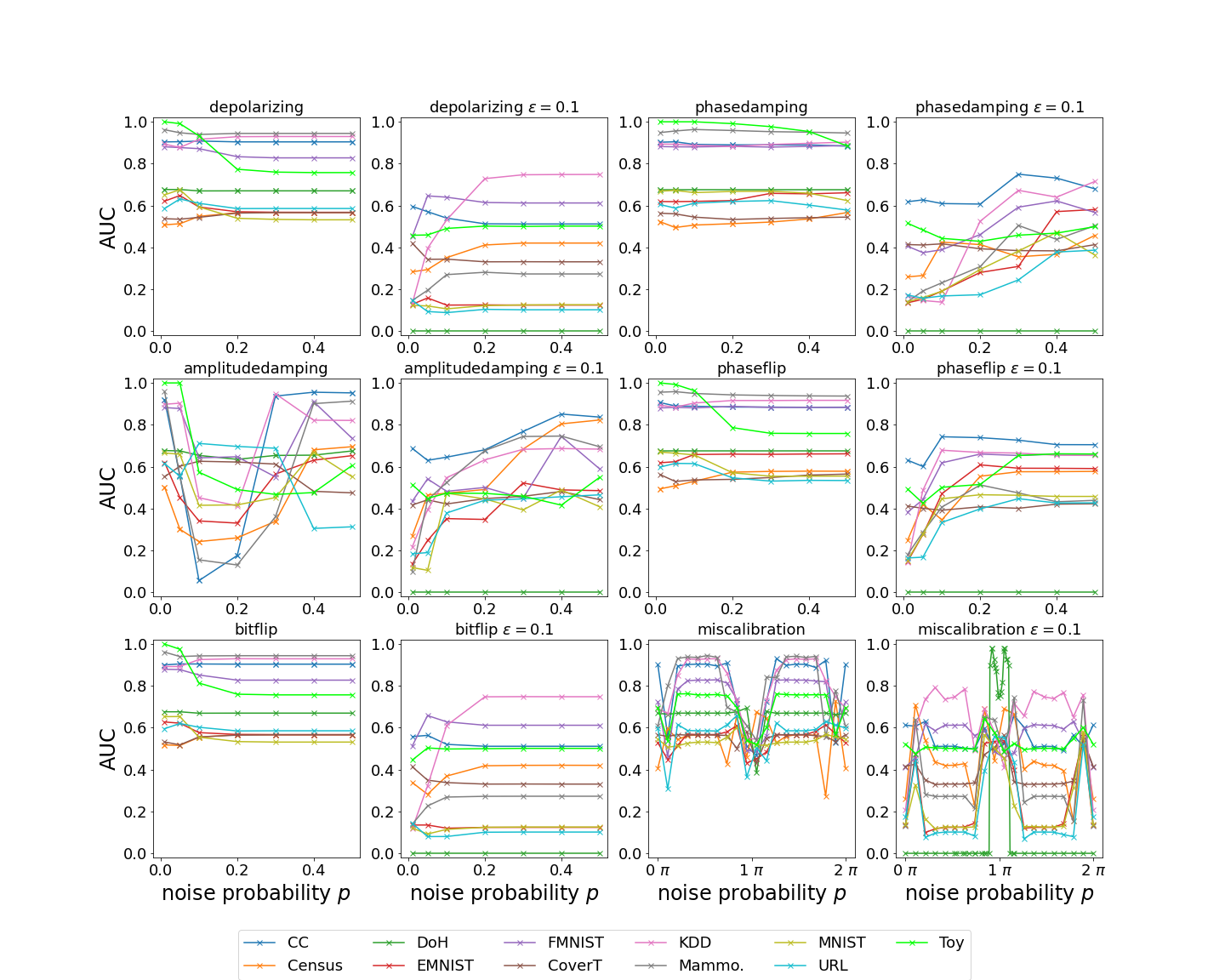}}
\caption{Left two columns: Influence of six different noise types on the QSVR. Right two columns: Influence of six different noise types on the QSVR for adversarial attacks of strength $\varepsilon=0.1$}
\label{fig_noise_channels}
\end{figure*}

The hardware results show that noise in NISQ devices usually affects the results obtained by the QSVR. Therefore, we examine the influence of six noise channels on the performance of the QSVR on eleven datasets in this section.
\cref{fig_noise_channels} presents the AUC of simulations across different noise levels for the channels described in \cref{Noisy Quantum Computing}, along with the impact of adversarial attacks of strength $\varepsilon = 0.1$ on these noisy simulations. The influence of noise on the adversarial robustness is investigated later, in \cref{Adversarial Robustness}.
The QSVR is largely robust against \emph{depolarizing}, \emph{phase damping}, \emph{phase flip}, and \emph{bit flip} noise, as the AUC for these noise types remains rather stable with increasing noise probability. A theoretical analysis of the robustness of quantum classifiers done by \cite{noise_robustness_LaRose} proves that single-qubit classifiers are robust against precisely these noise channels. 
Taking into account the findings of \cite{supervised_qml_are_kernel_schuld2021} that many short-term and fault-tolerant quantum models can be replaced by a general support vector machine whose kernel computes distances between data-encoding quantum states, we expect these results to transfer to our QSVR.

\emph{Amplitude damping} has a large effect on the model's behavior as the AUC decreases for all datasets at $p=0.1$ and $p=0.2$ except for CoverT and URL. At higher $p$, the AUC partially recovers and approaches roughly 0.5. 
These findings indicate that the QSVR is highly sensitive to amplitude damping, with performance under high noise degrading to that of a random classifier. 
Our observation is consistent with the analysis of \cite{noise_robustness_LaRose}, who showed that quantum classifiers lack robustness against this type of noise.

The curves of the models subject to \emph{miscalibration} noise show a periodicity of approximately $\pi$, with dips at about $k \pi$ for $k = 0,1,2$, and plateaus in between.
Small levels of miscalibration degrade the performance of the model, but when the noise level is above about $0.25 \pi$ , the AUC reaches a plateau at a level similar to the one for zero noise until the curve dips again around $p=\pi$.
The miscalibration introduces additional rotations in the circuit, effectively changing the states after feature encoding. This alters the fidelities and thereby impacts the reconstruction error and overall performance. 
We conclude that small degrees of miscalibration reduce model performance, and that this type of noise should be avoided in hardware.
However, we leave it open for future research to theoretically analyze the impact of miscalibration noise on the model.

\subsection{Adversarial Robustness} \label{Adversarial Robustness}
AD is often used in security-critical areas such as credit card fraud detection or network intrusion detection. Therefore, ML models used for AD must be robust to adversarial attacks.
For the following analysis, we note that for low-dimensional, tabular data sets, it is possible that a sample can be completely and effectively transformed into a sample of a different class at high attack strengths. In these cases, the effect on the AUC may be exaggerated and should be interpreted only as an upper bound on the performance drop.

\subsubsection{Noise-Free Adversarial Attacks} \label{Noise-Free Adversarial Attacks}
First, we consider the noise-free model and plot the obtained results of the PGD attacks up to a strength of $\varepsilon=0.5$ in \cref{fig_adv_attacks_noise_free}. 
The noise-free QSVR is highly vulnerable to adversarial attacks, as evidenced by the decrease in AUC for small attack strengths of $\varepsilon=0.01$ for all datasets except Toy. The largest decrease occurs for the DoH dataset, whose AUC drops by an order of magnitude from 0.67 to 0.06 for the $\varepsilon=0.01$ attack. As the attack strength increases, the AUC continues to decrease for all datasets until it approaches 0.0 for $p=0.3$. 
We conclude that techniques to increase the adversarial robustness of the model should be investigated.

The unchanged AUC of Toy for $\varepsilon=0.01$ can be explained by the creation process of the dataset. The dataset was created to be linearly separable with a separation distance of 0.4 between the classes, so the data points must be shifted a large distance in feature space to be misclassified. However, since the AUC drops to 0.0 for $\varepsilon=0.3$, which is smaller than the separation distance, we conclude that the QSVR is susceptible to overfitting.

\subsubsection{Noisy Adversarial Attacks} \label{Noisy Adversarial Attacks}
\begin{figure}[htbp]
\centerline{\includegraphics[width=0.5\textwidth]{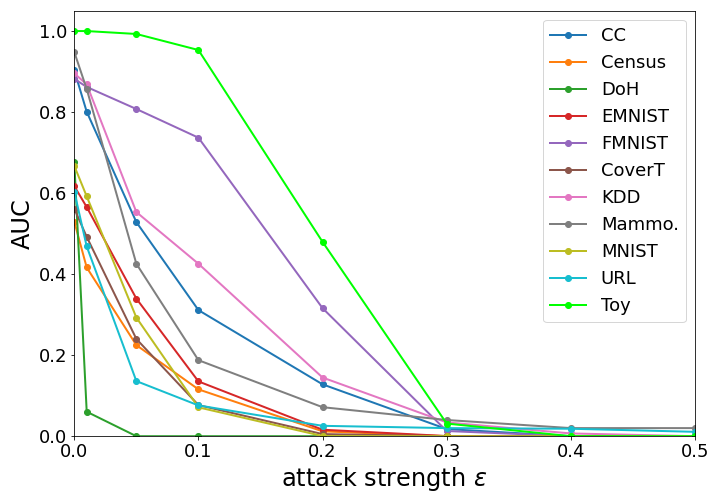}}
\caption{Influence of adversarial attacks on the noise-free models.}
\label{fig_adv_attacks_noise_free}
\end{figure}

Second, the performance of the noisy QSVR when subjected to adversarial perturbations of strength $\varepsilon=0.1$ is shown in \cref{fig_noise_channels}. Omitting \emph{miscalibration}, we find that for noise levels below $p=0.1$, the AUC is low for most noise types and datasets, following the trend of high adversarial vulnerability observed in \cref{fig_adv_attacks_noise_free}. At higher noise levels, however, the AUC typically recovers to some extent, reaching a plateau at about 0.5. This indicates that the model transitions to a random classifier, showing that the adversarial attacks are so powerful that quantum noise cannot improve performance beyond that of a random classifier. 
Other researchers report similar results, noting that random noise and adversarial noise are fundamentally different, and that even models resilient to random noise are often still vulnerable to adversarial noise \cite{fawzi2016robustnessclassifiersadversarialrandom}.

\emph{Miscalibration} noise influences both attacked and unattacked models in a similar manner. For models under attack, the plateaus occur at lower AUC values, while near $p = k\pi$ for $k = 0,1,2$ the AUC often rises above the adjacent plateaus, producing spikes rather than drops. Such spikes exceed both the AUC at $p=0$ and the 0.5 threshold only for DoH, FMNIST, KDD, Toy, Census, and Mammo., indicating that small amounts of this noise can enhance performance under attack. As the remaining datasets do not exhibit such gains, we conclude that miscalibration noise cannot be regarded as a reliable means of enhancing adversarial robustness.

The DoH dataset is an outlier, with an AUC of 0.0 across almost all noise types and strengths, which is explained by its extreme vulnerability to adversarial attacks seen in \cref{fig_adv_attacks_noise_free}. 
This vulnerability can be explained by \cref{app:fig:dataset_analysis} in the appendix, showing the minimum p-value from the \emph{Kolmogorov–Smirnov} test and the maximum feature variance within the test set. The DoH dataset has a medium p-value ($\approx 10^{-8}$) combined with a very low variance ($\approx2\times10^{-2}$). The p-value indicates the probability that the normal and anomalous samples originate from the same distribution, while the very low variance suggests a high degree of similarity between all samples, especially between the normal and anomalous data. As a result, the DoH dataset is difficult to classify, and even tiny adversarial attacks of strength $\varepsilon = 0.01$ lead to manipulations about five times greater than the variance within the dataset. 

Notably, the AUC for the DoH dataset with \emph{miscalibration} noise is $0.0$ across nearly all noise levels and peaks only around $p=\pi$, where it approaches $1.0$. 
An analysis of the adversarial test kernels for DoH under \emph{miscalibration} noise (\cref{tab:DoH_kernel}) highlights major differences between a high-performing run ($p = 2.9 \approx 0.9 \pi$, $\text{AUC} = 0.98$) and a low-performing run ($p = 1.7 \approx 0.5 \pi$, $\text{AUC} = 0.00$). The mean kernel values in the high-AUC case are four orders of magnitude larger than in the low-AUC case, and hence the disparity between classes is more pronounced, enabling the SVR to separate them more effectively. Since the kernel entries represent the fidelity of the states created by embedding two samples, we conclude that \emph{miscalibration} noise close to $\pi$ increases within-class similarity and thus elevates the kernel values.

In light of this, the overrotation introduced by \emph{miscalibration} noise can be interpreted as the addition of fixed-parameter rotation gates independent of the input data. As these parameters strongly affect model performance, this underscores the importance of employing dataset-specific kernels, such as trainable kernels.

\begin{table}[htbp]
\centering
\caption{Analysis of the adversarial test kernel for the DoH dataset subject to miscalibration noise.}  \label{tab:DoH_kernel}
\begin{tabular}{rrrrr}
\toprule
p & AUC & class & mean kernel value \\\midrule
\multirow[t]{2}{*}{2.9} & \multirow[t]{2}{*}{0.98} & 0 & 1.743e-01 $\pm$ 5.157e-02 \\
 &  & 1 & 1.734e-01 $\pm$ 5.121e-02 \\
\multirow[t]{2}{*}{1.7} & \multirow[t]{2}{*}{0.00} & 0 & 1.800e-05 $\pm$ 7.360e-06 \\
 &  & 1 & 1.796e-05 $\pm$ 7.395e-06 \\
\bottomrule
\end{tabular}

\end{table}

The models attacked with $\varepsilon=0.01$ and $\varepsilon=0.3$  do not provide new insights, as they exhibit similar behavior to the $\varepsilon=0.1$ attacks above, and can be obtained from the authors upon reasonable request.

We conclude that quantum noise is not suited for increasing the adversarial robustness of the QSVR. This finding is consistent with prior research \cite{noise_influence_winderl}, where the authors suggest that adding noise to QML models to increase the adversarial robustness is unlikely to be beneficial in practice.

\subsubsection{Adversarial Training}
Adversarial training is a straightforward approach to increasing the adversarial robustness of supervised learning algorithms. 
\cref{tab:adv_training_auc} reveals that adversarial training increases the AUC for the adversarial test set on seven out of eleven datasets, and the average AUC over all datasets rises from 0.28 to 0.31. However, the increase in AUC is small, and except for FMNIST and Toy, the AUC remains below 0.5. For the test set without adversarial samples, the AUC decreases through adversarial training on six out of eleven datasets, and the average declines from 0.75 to 0.71.
\cref{tab:adv_training_ratios} shows the ratio of correctly classified normal samples to the total number of normal samples, as well as the same ratio for the anomalies. For normal data, the ratio is $\frac{\text{tn}}{\text{tn} + \text{fp}}$ , and for the anomalies it is $\frac{\text{tp}}{\text{tp} + \text{fn}}$. 
We observe that retraining increases the classification ratio for the normal data of the Toy dataset from 0.78 to 0.94, while the ratio for the anomalies remains unchanged at 1.00. For KDD, we report similar results, but the increase in the classification ratio of the normal data through retraining is smaller. 
This shows that adversarial retraining can lead to more normal samples being classified correctly without influencing the classification of the anomalies, since the latter are not contained in the training set. However, this was not observed for other datasets and was most pronounced for the synthetic dataset, suggesting this behavior requires a large separation distance between the two classes.

We conclude that adversarial training cannot be used to reliably harden the QSVR against adversarial attacks. We attribute this to the \emph{semisupervised} setting, meaning that only normal samples are available during training. 

\begin{table}[htbp]
\centering
\caption{AUCs of the models for the test set and adversarial test set with and without adversarial training.}  \label{tab:adv_training_auc}
\begin{tabular}{p{1.0cm}p{1.1cm}<{\raggedright}p{1.1cm}<{\raggedright}p{1.1cm}<{\raggedright}p{1.1cm}<{\raggedright}}
\toprule
Dataset & Test AUC w/ retraining & Test AUC w/o retraining & Adv AUC w/ retraining & Adv AUC w/o retraining \\
\midrule
CC & 0.85 & 0.90 & 0.37 & 0.31 \\
Census & 0.62 & 0.53 & 0.05 & 0.12 \\
DoH & 0.68 & 0.68 & 0.00 & 0.00 \\
EMNIST & 0.63 & 0.62 & 0.17 & 0.14 \\
FMNIST & 0.91 & 0.88 & 0.73 & 0.74 \\
CoverT & 0.44 & 0.56 & 0.18 & 0.08 \\
KDD & 0.71 & 0.90 & 0.46 & 0.43 \\
Mammo. & 0.74 & 0.95 & 0.22 & 0.19 \\
MNIST & 0.62 & 0.67 & 0.11 & 0.07 \\
URL & 0.59 & 0.60 & 0.08 & 0.08 \\
Toy & 1.00 & 1.00 & 1.00 & 0.95 \\
\textbf{Mean} & \textbf{0.71} & \textbf{0.75} & \textbf{0.31} & \textbf{0.28} \\
\bottomrule
\end{tabular}

\end{table}

\begin{table}[htbp]
\centering
\caption{Ratios of correctly classified adversarial normal and anomalous samples.}  \label{tab:adv_training_ratios}
\begin{tabular}{p{1.0cm}p{0.5cm}p{0.5cm}p{0.5cm}p{0.5cm}p{0.5cm}p{0.5cm}p{0.5cm}p{0.5cm}}
\toprule
Dataset & \multicolumn{2}{c}{retraining} & \multicolumn{2}{c}{no retraining} \\
 & norm. & anom. & norm. & anom. \\
\midrule
CC & 0.96 & 0.28 & 0.98 & 0.24 \\
Census & 1.00 & 0.00 & 1.00 & 0.00 \\
DoH & 1.00 & 0.00 & 1.00 & 0.00 \\
EMNIST & 1.00 & 0.02 & 0.98 & 0.02 \\
FMNIST & 0.86 & 0.66 & 0.86 & 0.68 \\
CoverT & 1.00 & 0.00 & 1.00 & 0.00 \\
KDD & 1.00 & 0.34 & 0.96 & 0.34 \\
Mammo. & 0.96 & 0.18 & 1.00 & 0.14 \\
MNIST & 1.00 & 0.00 & 1.00 & 0.00 \\
URL & 0.96 & 0.08 & 0.98 & 0.04 \\
Toy & 0.94 & 1.00 & 0.78 & 1.00 \\
\textbf{Mean} & \textbf{0.97} & \textbf{0.23} & \textbf{0.96} & \textbf{0.22} \\
\bottomrule
\end{tabular}

\end{table}

\section{\uppercase{Conclusion and Outlook}} \label{Conclusion and Outlook}
We first benchmarked our QSVR for semisupervised AD on 27-qubits IBM hardware and found that the average AUC was slightly lower than that of the noiseless simulation (0.72 compared to 0.76). However, the QSVR outperformed the noiseless simulation on two out of eleven datasets. 

Second, the influence of six noise channels on the performance of the QSVR was evaluated, revealing that it is largely robust against \emph{deporarizing}, \emph{phasedamping}, \emph{phase flip} and \emph{bit flip} noise.
\emph{Amplitude damping}, on the other hand, results in the most significant degradation of the model and \emph{miscalibration} noise also impacts performance.

Finally, the adversarial robustness of the (noisy) QSVR was assessed, and it was observed that the model is highly vulnerable to adversarial attacks. Even weak PGD attacks with a strength of $\varepsilon=0.01$ can reduce the AUC by up to an order of magnitude. 
Introducing quantum noise does not yield any beneficial effect, neither on the unattacked model's performance nor on its adversarial robustness.
Moreover, adversarial training does not reliably improve the adversarial robustness of the model.
Consequently, we conclude that the QSVR demonstrates potential for semisupervised AD in the NISQ era, however, special attention should be paid to the vulnerability to adversarial attacks and \emph{amplitude damping} and \emph{miscalibration} noise.

We emphasize the importance of employing dataset-specific kernels and recommend exploring \emph{trainable kernels} to further enhance the performance of QML models.
In addition, a systematic study is needed to understand why, on certain datasets, the hardware-executed QSVR surpasses its simulated counterpart.
Furthermore, a theoretical analysis of the effects of \emph{miscalibration} noise and its influence on kernel values and reconstruction errors will also be helpful.
Beyond these directions, future work should explore scaling the model to larger numbers of qubits and, critically, developing \emph{effective defenses against adversarial attacks} to strengthen the robustness of QSVR in security-sensitive applications.

\section*{\uppercase{Acknowledgements}}
The research is part of the Munich Quantum Valley, which is supported by the Bavarian state government with funds from the Hightech Agenda Bayern Plus.

\bibliographystyle{apalike}
{\small
\bibliography{qad}}

@article{J_ger_2023,
   title={Universal expressiveness of variational quantum classifiers and quantum kernels for support vector machines},
   volume={14},
   ISSN={2041-1723},
   url={http://dx.doi.org/10.1038/s41467-023-36144-5},
   DOI={10.1038/s41467-023-36144-5},
   number={1},
   journal={Nature Communications},
   publisher={Springer Science and Business Media LLC},
   author={Jäger, Jonas and Krems, Roman V.},
   year={2023},
   month=feb,
}

@misc{fawzi2016robustnessclassifiersadversarialrandom,
      title={Robustness of classifiers: from adversarial to random noise}, 
      author={Alhussein Fawzi and Seyed-Mohsen Moosavi-Dezfooli and Pascal Frossard},
      year={2016},
      eprint={1608.08967},
      archivePrefix={arXiv},
      primaryClass={cs.LG},
      url={https://arxiv.org/abs/1608.08967}, 
}

@Article{Domingo2023,
author={Domingo, L.
and Carlo, G.
and Borondo, F.},
title={Taking advantage of noise in quantum reservoir computing},
journal={Scientific Reports},
year={2023},
day={31},
volume={13},
number={1},
pages={8790},
issn={2045-2322},
doi={10.1038/s41598-023-35461-5},
url={https://doi.org/10.1038/s41598-023-35461-5}
}

@inbook{Escudero_2023,
   title={Assessing the Impact of Noise on Quantum Neural Networks: An Experimental Analysis},
   ISBN={9783031407253},
   ISSN={1611-3349},
   url={http://dx.doi.org/10.1007/978-3-031-40725-3_27},
   DOI={10.1007/978-3-031-40725-3_27},
   booktitle={Hybrid Artificial Intelligent Systems},
   publisher={Springer Nature Switzerland},
   author={Escudero, Erik Terres and Alamo, Danel Arias and Gómez, Oier Mentxaka and Bringas, Pablo García},
   year={2023},
   pages={314–325} }

@inproceedings{denoising_autoencoder,
author = {Vincent, Pascal and Larochelle, Hugo and Bengio, Yoshua and Manzagol, Pierre-Antoine},
title = {Extracting and composing robust features with denoising autoencoders},
year = {2008},
isbn = {9781605582054},
publisher = {Association for Computing Machinery},
address = {New York, NY, USA},
url = {https://doi.org/10.1145/1390156.1390294},
doi = {10.1145/1390156.1390294},
booktitle = {Proceedings of the 25th International Conference on Machine Learning},
pages = {1096–1103},
numpages = {8},
location = {Helsinki, Finland},
series = {ICML '08}
}

@ARTICLE{Ruff2020,
  author={Ruff, Lukas and Kauffmann, Jacob R. and Vandermeulen, Robert A. and Montavon, Grégoire and Samek, Wojciech and Kloft, Marius and Dietterich, Thomas G. and Müller, Klaus-Robert},
  journal={Proceedings of the IEEE}, 
  title={A Unifying Review of Deep and Shallow Anomaly Detection}, 
  year={2021},
  volume={109},
  number={5},
  pages={756-795},
  doi={10.1109/JPROC.2021.3052449}}

@article{quantum_kernel_advantage_discrete_log_2021,
   author = {Yunchao Liu and Srinivasan Arunachalam and Kristan Temme},
   doi = {10.1038/s41567-021-01287-z},
   issn = {17452481},
   issue = {9},
   journal = {Nature Physics},
   month = {9},
   pages = {1013-1017},
   publisher = {Nature Research},
   title = {A rigorous and robust quantum speed-up in supervised machine learning},
   volume = {17},
   year = {2021},
}

@article{quantum_noise_against_adv_attacks_Du,
  title = {Quantum noise protects quantum classifiers against adversaries},
  author = {Du, Yuxuan and Hsieh, Min-Hsiu and Liu, Tongliang and Tao, Dacheng and Liu, Nana},
  journal = {Phys. Rev. Res.},
  volume = {3},
  issue = {2},
  pages = {023153},
  numpages = {18},
  year = {2021},
  month = {05},
  publisher = {American Physical Society},
  doi = {10.1103/PhysRevResearch.3.023153},
}

@article{noise_improve_adv_robustness_Huang,
doi = {10.1088/1367-2630/ace8b4},
year = {2023},
month = {08},
publisher = {IOP Publishing},
volume = {25},
number = {8},
pages = {083019},
author = {Chenyi Huang and Shibin Zhang},
title = {Enhancing adversarial robustness of quantum neural networks by adding noise layers},
journal = {New Journal of Physics},
}

@InProceedings{URL,
author="Mamun, Mohammad Saiful Islam
and Rathore, Mohammad Ahmad
and Lashkari, Arash Habibi
and Stakhanova, Natalia
and Ghorbani, Ali A.",
editor="Chen, Jiageng
and Piuri, Vincenzo
and Su, Chunhua
and Yung, Moti",
title="Detecting Malicious URLs Using Lexical Analysis",
booktitle="Network and System Security",
year="2016",
publisher="Springer International Publishing",
address="Cham",
pages="467--482",
isbn="978-3-319-46298-1"
}

@article{mammo,
author = {Woods, KEVIN S. and Doss, CHRISTOPHER C. and Bowyer, KEVIN W. and Solka, JEFFREY L. and Priebe, CAREY E. and Kegelmeyer, W. PHILIP},
title = {COMPARATIVE EVALUATION OF PATTERN RECOGNITION TECHNIQUES FOR DETECTION OF MICROCALCIFICATIONS IN MAMMOGRAPHY},
journal = {International Journal of Pattern Recognition and Artificial Intelligence},
volume = {07},
number = {06},
pages = {1417-1436},
year = {1993},
doi = {10.1142/S0218001493000698},
eprint = {https://doi.org/10.1142/S0218001493000698},
}

@ARTICLE{mnist,
  author={Lecun, Y. and Bottou, L. and Bengio, Y. and Haffner, P.},
  journal={Proceedings of the IEEE}, 
  title={Gradient-based learning applied to document recognition}, 
  year={1998},
  volume={86},
  number={11},
  pages={2278-2324},
  doi={10.1109/5.726791}}

@INPROCEEDINGS{KDD,
  author={Tavallaee, Mahbod and Bagheri, Ebrahim and Lu, Wei and Ghorbani, Ali A.},
  booktitle={2009 IEEE Symposium on Computational Intelligence for Security and Defense Applications}, 
  title={A detailed analysis of the KDD CUP 99 data set}, 
  year={2009},
  volume={},
  number={},
  pages={1-6},
  doi={10.1109/CISDA.2009.5356528}}

@misc{fmnist,
      title={Fashion-MNIST: a Novel Image Dataset for Benchmarking Machine Learning Algorithms}, 
      author={Han Xiao and Kashif Rasul and Roland Vollgraf},
      year={2017},
      eprint={1708.07747},
      archivePrefix={arXiv},
      primaryClass={cs.LG}
}

@INPROCEEDINGS{emnist,
  author={Cohen, Gregory and Afshar, Saeed and Tapson, Jonathan and van Schaik, André},
  booktitle={2017 International Joint Conference on Neural Networks (IJCNN)}, 
  title={EMNIST: Extending MNIST to handwritten letters}, 
  year={2017},
  volume={},
  number={},
  pages={2921-2926},
  doi={10.1109/IJCNN.2017.7966217}}

@INPROCEEDINGS{DoH,
  author={MontazeriShatoori, Mohammadreza and Davidson, Logan and Kaur, Gurdip and Habibi Lashkari, Arash},
  booktitle={2020 IEEE Intl Conf on Dependable, Autonomic and Secure Computing, Intl Conf on Pervasive Intelligence and Computing, Intl Conf on Cloud and Big Data Computing, Intl Conf on Cyber Science and Technology Congress (DASC/PiCom/CBDCom/CyberSciTech)}, 
  title={Detection of DoH Tunnels using Time-series Classification of Encrypted Traffic}, 
  year={2020},
  volume={},
  number={},
  pages={63-70},
  doi={10.1109/DASC-PICom-CBDCom-CyberSciTech49142.2020.00026}}

@article{forest_type,
title = {Comparative accuracies of artificial neural networks and discriminant analysis in predicting forest cover types from cartographic variables},
journal = {Computers and Electronics in Agriculture},
volume = {24},
number = {3},
pages = {131-151},
year = {1999},
issn = {0168-1699},
doi = {https://doi.org/10.1016/S0168-1699(99)00046-0},
author = {Jock A. Blackard and Denis J. Dean},
}

@misc{census_income,
author = "Dua, Dheeru and Graff, Casey",
year = "2017",
title = "{UCI} Machine Learning Repository",
url = "http://archive.ics.uci.edu/ml",
institution = "University of California, Irvine, School of Information and Computer Sciences" }

@INPROCEEDINGS{cc,
  author={Pozzolo, Andrea Dal and Caelen, Olivier and Johnson, Reid A. and Bontempi, Gianluca},
  booktitle={2015 IEEE Symposium Series on Computational Intelligence}, 
  title={Calibrating Probability with Undersampling for Unbalanced Classification}, 
  year={2015},
  volume={},
  number={},
  pages={159-166},
  doi={10.1109/SSCI.2015.33}}

@misc{pennylane,
      title={PennyLane: Automatic differentiation of hybrid quantum-classical computations}, 
      author={Ville Bergholm and Josh Izaac and Maria Schuld and Christian Gogolin and Shahnawaz Ahmed and Vishnu Ajith and M. Sohaib Alam and Guillermo Alonso-Linaje and B. AkashNarayanan and Ali Asadi and Juan Miguel Arrazola and Utkarsh Azad and Sam Banning and Carsten Blank and Thomas R Bromley and Benjamin A. Cordier and Jack Ceroni and Alain Delgado and Olivia Di Matteo and Amintor Dusko and Tanya Garg and Diego Guala and Anthony Hayes and Ryan Hill and Aroosa Ijaz and Theodor Isacsson and David Ittah and Soran Jahangiri and Prateek Jain and Edward Jiang and Ankit Khandelwal and Korbinian Kottmann and Robert A. Lang and Christina Lee and Thomas Loke and Angus Lowe and Keri McKiernan and Johannes Jakob Meyer and J. A. Montañez-Barrera and Romain Moyard and Zeyue Niu and Lee James O'Riordan and Steven Oud and Ashish Panigrahi and Chae-Yeun Park and Daniel Polatajko and Nicolás Quesada and Chase Roberts and Nahum Sá and Isidor Schoch and Borun Shi and Shuli Shu and Sukin Sim and Arshpreet Singh and Ingrid Strandberg and Jay Soni and Antal Száva and Slimane Thabet and Rodrigo A. Vargas-Hernández and Trevor Vincent and Nicola Vitucci and Maurice Weber and David Wierichs and Roeland Wiersema and Moritz Willmann and Vincent Wong and Shaoming Zhang and Nathan Killoran},
      year={2022},
      eprint={1811.04968},
      archivePrefix={arXiv},
      primaryClass={quant-ph},
}

@misc{qiskit,
      title={Quantum computing with {Q}iskit},
      author={Javadi-Abhari, Ali and Treinish, Matthew and Krsulich, Kevin and Wood, Christopher J. and Lishman, Jake and Gacon, Julien and Martiel, Simon and Nation, Paul D. and Bishop, Lev S. and Cross, Andrew W. and Johnson, Blake R. and Gambetta, Jay M.},
      year={2024},
      doi={10.48550/arXiv.2405.08810},
      eprint={2405.08810},
      archivePrefix={arXiv},
      primaryClass={quant-ph}
}

@misc{noise_influence_winderl,
      title={Quantum Neural Networks under Depolarization Noise: Exploring White-Box Attacks and Defenses}, 
      author={David Winderl and Nicola Franco and Jeanette Miriam Lorenz},
      year={2023},
      eprint={2311.17458},
      archivePrefix={arXiv},
      primaryClass={quant-ph},
}

@InProceedings{noise_robustness_Yao,
  title = 	 {Noise-Robust End-to-End Quantum Control using Deep Autoregressive Policy Networks},
  author =       {Yao, Jiahao and Kottering, Paul and Gundlach, Hans and Lin, Lin and Bukov, Marin},
  booktitle = 	 {Proceedings of the 2nd Mathematical and Scientific Machine Learning Conference},
  pages = 	 {1044--1081},
  year = 	 {2022},
  editor = 	 {Bruna, Joan and Hesthaven, Jan and Zdeborova, Lenka},
  volume = 	 {145},
  series = 	 {Proceedings of Machine Learning Research},
  month = 	 {08},
  publisher =    {PMLR},
  url = 	 {https://proceedings.mlr.press/v145/yao22a.html},
}

@article{noise_robustness_LaRose,
  title = {Robust data encodings for quantum classifiers},
  author = {LaRose, Ryan and Coyle, Brian},
  journal = {Phys. Rev. A},
  volume = {102},
  issue = {3},
  pages = {032420},
  numpages = {24},
  year = {2020},
  month = {09},
  publisher = {American Physical Society},
  doi = {10.1103/PhysRevA.102.032420},
}

@ARTICLE{noise_robustness_Nguyen,
  title    = "Quantum learning with noise and decoherence: a robust quantum
              neural network",
  author   = "Nguyen, Nam H and Behrman, Elizabeth C and Steck, James E",
  journal  = "Quantum Machine Intelligence",
  volume   =  2,
  number   =  1,
  pages    = "1",
  month    =  jan,
  year     =  2020
}

@misc{noise_beneficial_winderl,
      title={Constructing Optimal Noise Channels for Enhanced Robustness in Quantum Machine Learning}, 
      author={David Winderl and Nicola Franco and Jeanette Miriam Lorenz},
      year={2024},
      eprint={2404.16417},
      archivePrefix={arXiv},
      primaryClass={quant-ph},
      url={https://arxiv.org/abs/2404.16417}, 
}

@INPROCEEDINGS{noise_beneficial_ju,
  author={Ju, Keyi and Qin, Xiaoqi and Zhong, Hui and Zhang, Xinyue and Pan, Miao and Liu, Baoling},
  booktitle={ICC 2024 - IEEE International Conference on Communications}, 
  title={Harnessing Inherent Noises for Privacy Preservation in Quantum Machine Learning}, 
  year={2024},
  volume={},
  number={},
  pages={1121-1126},
  doi={10.1109/ICC51166.2024.10622663}}

@article{noise_influence_Diego,
  title = {Effects of noise on the overparametrization of quantum neural networks},
  author = {Garc\'{\i}a-Mart\'{\i}n, Diego and Larocca, Mart\'{\i}n and Cerezo, M.},
  journal = {Phys. Rev. Res.},
  volume = {6},
  issue = {1},
  pages = {013295},
  numpages = {17},
  year = {2024},
  publisher = {American Physical Society},
  doi = {10.1103/PhysRevResearch.6.013295},
  url = {https://link.aps.org/doi/10.1103/PhysRevResearch.6.013295}
}

@misc{quantum_one_class_SVM,
      title={Unsupervised quantum machine learning for fraud detection}, 
      author={Oleksandr Kyriienko and Einar B. Magnusson},
      year={2022},
      eprint={2208.01203},
      archivePrefix={arXiv},
      primaryClass={quant-ph}
}

@INPROCEEDINGS{Mafu2021,
  author={Mafu, Mhlambululi and Senekane, Makhamisa},
  booktitle={2021 International Conference on Electrical, Computer and Energy Technologies (ICECET)}, 
  title={Design and Implementation of Efficient Quantum Support Vector Machine}, 
  year={2021},
  volume={},
  number={},
  pages={1-4},
  doi={10.1109/ICECET52533.2021.9698509}}

@INPROCEEDINGS{Ahmad2021,
  author={Farhan Ahmad, Syed and Rawat, Raghav and Moharir, Minal},
  booktitle={2021 International Conference on Computational Intelligence and Knowledge Economy (ICCIKE)}, 
  title={Quantum Machine Learning with HQC Architectures using non-Classically Simulable Feature Maps}, 
  year={2021},
  volume={},
  number={},
  pages={345-349},
  doi={10.1109/ICCIKE51210.2021.9410753}}

@INPROCEEDINGS{Delilbasic2021,
  author={Delilbasic, Amer and Cavallaro, Gabriele and Willsch, Madita and Melgani, Farid and Riedel, Morris and Michielsen, Kristel},
  booktitle={2021 IEEE International Geoscience and Remote Sensing Symposium IGARSS}, 
  title={Quantum Support Vector Machine Algorithms for Remote Sensing Data Classification}, 
  year={2021},
  volume={},
  number={},
  pages={2608-2611},
  doi={10.1109/IGARSS47720.2021.9554802}}

@ARTICLE{Havlicek2019,
  title    = "Supervised learning with quantum-enhanced feature spaces",
  author   = "Havl{\'\i}{\v c}ek, Vojt{\v e}ch and C{\'o}rcoles, Antonio D and
              Temme, Kristan and Harrow, Aram W and Kandala, Abhinav and Chow,
              Jerry M and Gambetta, Jay M",
  journal  = "Nature",
  volume   =  567,
  number   =  7747,
  pages    = "209--212",
  year     =  2019
}

@misc{supervised_qml_are_kernel_schuld2021,
      title={Supervised quantum machine learning models are kernel methods}, 
      author={Maria Schuld},
      year={2021},
      eprint={2101.11020},
      archivePrefix={arXiv},
      primaryClass={quant-ph},
}

@misc{adv_robustness_wendlinger,
      title={A Comparative Analysis of Adversarial Robustness for Quantum and Classical Machine Learning Models}, 
      author={Maximilian Wendlinger and Kilian Tscharke and Pascal Debus},
      year={2024},
      eprint={2404.16154},
      archivePrefix={arXiv},
      primaryClass={cs.LG},
}

@misc{adv_attacks_goodfellow,
      title={Explaining and Harnessing Adversarial Examples}, 
      author={Ian J. Goodfellow and Jonathon Shlens and Christian Szegedy},
      year={2015},
      eprint={1412.6572},
      archivePrefix={arXiv},
      primaryClass={stat.ML},
      url={https://arxiv.org/abs/1412.6572}, 
}

@misc{pgd_madry,
      title={Towards Deep Learning Models Resistant to Adversarial Attacks}, 
      author={Aleksander Madry and Aleksandar Makelov and Ludwig Schmidt and Dimitris Tsipras and Adrian Vladu},
      year={2019},
      eprint={1706.06083},
      archivePrefix={arXiv},
      primaryClass={stat.ML},
}

@misc{adv_attacks_szegedy,
      title={Intriguing properties of neural networks}, 
      author={Christian Szegedy and Wojciech Zaremba and Ilya Sutskever and Joan Bruna and Dumitru Erhan and Ian Goodfellow and Rob Fergus},
      year={2014},
      eprint={1312.6199},
      archivePrefix={arXiv},
      primaryClass={cs.CV},
}

@book{Schuld_ML_w_QC,
	address = {Cham},
	series = {Quantum {Science} and {Technology}},
	title = {Machine {Learning} with {Quantum} {Computers}},
	copyright = {https://www.springer.com/tdm},
	isbn = {978-3-030-83097-7},
	language = {en},
	publisher = {Springer International Publishing},
	author = {Schuld, Maria and Petruccione, Francesco},
	year = {2021},
	doi = {10.1007/978-3-030-83098-4},
}

@article{Schuld_QML_in_Feature_Hilbert_Spaces,
  title = {Quantum Machine Learning in Feature Hilbert Spaces},
  author = {Schuld, Maria and Killoran, Nathan},
  journal = {Phys. Rev. Lett.},
  volume = {122},
  issue = {4},
  pages = {040504},
  numpages = {6},
  year = {2019},
  month = {02},
  publisher = {American Physical Society},
  doi = {10.1103/PhysRevLett.122.040504},
}

@misc{preskill_lectures_noise,
	title = {Lecture {Notes} for {Ph219}/{CS219}: {Quantum} {Information} {Chapter} 3},
	language = {en},
	author = {Preskill, John},
    month = {10},
    year = {2018},
    publisher = {California Institute of Technology},
}

@article{Wallman_Noise,
doi = {10.1088/1367-2630/17/11/113020},
year = {2015},
month = {11},
publisher = {IOP Publishing},
volume = {17},
number = {11},
pages = {113020},
author = {Joel Wallman and Chris Granade and Robin Harper and Steven T Flammia},
title = {Estimating the coherence of noise},
journal = {New Journal of Physics},
}

@article{Suter_Noise,
  title = {Colloquium: Protecting quantum information against environmental noise},
  author = {Suter, Dieter and \'Alvarez, Gonzalo A.},
  journal = {Rev. Mod. Phys.},
  volume = {88},
  issue = {4},
  pages = {041001},
  numpages = {23},
  year = {2016},
  month = {10},
  publisher = {American Physical Society},
  doi = {10.1103/RevModPhys.88.041001},
}

@misc{kaufmann2023_noise,
      title={Characterization of Coherent Errors in Noisy Quantum Devices}, 
      author={Noah Kaufmann and Ivan Rojkov and Florentin Reiter},
      year={2023},
      eprint={2307.08741},
      archivePrefix={arXiv},
      primaryClass={quant-ph},
}

@book{Nielsen, 
place={Cambridge}, 
title={Quantum Computation and Quantum Information: 10th Anniversary Edition}, publisher={Cambridge University Press}, 
author={Nielsen, Michael A. and Chuang, Isaac L.}, 
year={2010}
}

@INPROCEEDINGS {Tscharke,
author = {K. Tscharke and S. Issel and P. Debus},
booktitle = {2023 IEEE International Conference on Quantum Computing and Engineering (QCE)},
title = {Semisupervised Anomaly Detection using Support Vector Regression with Quantum Kernel},
year = {2023},
volume = {},
issn = {},
pages = {611-620},
doi = {10.1109/QCE57702.2023.00075},
publisher = {IEEE Computer Society},
address = {Los Alamitos, CA, USA},
month = {09}
}

@article{Kottmann2021,
  title = {Variational quantum anomaly detection: Unsupervised mapping of phase diagrams on a physical quantum computer},
  author = {Kottmann, Korbinian and Metz, Friederike and Fraxanet, Joana and Baldelli, Niccol\`o},
  journal = {Phys. Rev. Res.},
  volume = {3},
  issue = {4},
  pages = {043184},
  numpages = {9},
  year = {2021},
  month = {12},
  publisher = {American Physical Society},
  doi = {10.1103/PhysRevResearch.3.043184},
}

\section*{\uppercase{Appendix}}
\begin{table*}[!b]
\small
\centering
\caption{Overview of the datasets used for the experiments.}  \label{app:tab:overview_datasets}
\begin{tabular}{cccc}
\toprule 
Dataset & Reference & Normal class & Anomalous class \\
\midrule 
CC & \cite{cc} & Normal & Anomalous \\
Census & \cite{census_income} & $\leq$50k & $>$50k \\
CoverT & \cite{forest_type} & 1-4 & 5-7 \\
DoH & \cite{DoH} & Benign & Malicious \\
EMNIST & \cite{emnist} & A-M & N-Z \\
FMNIST & \cite{fmnist} & 0-4 & 5-9 \\
KDD & \cite{KDD} & Normal & Anomalous \\
MNIST & \cite{mnist} & 0-4 & 5-9 \\
Mammo & \cite{mammo} & Normal & Malignant \\
Toy & / & Normal & Anomalous \\
URL & \cite{URL} & Benign & Non-benign \\
\bottomrule 
\end{tabular}

\end{table*}

\begin{figure}[htb]
\centerline{\includegraphics[width=0.5\textwidth]{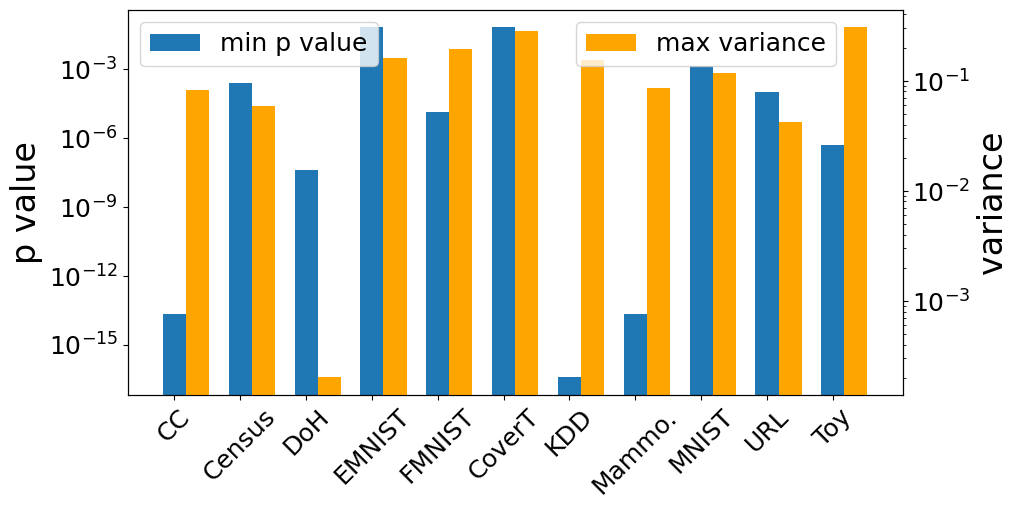}}
\caption{Left axis: min. p-value obtained from the Kolmogorov-Smirnov test. It gives the probability of the normal and anomalous samples being from the same distribution. Right axis: max. variance within the test set. 
Both values are build feature-wise and then the min/max value is plotted.}
\label{app:fig:dataset_analysis}
\end{figure}

\end{document}